\documentstyle[12pt,twoside,fleqn,epsfig,espcrc1]{article} 
\begin{document}

\title{To bind or not to bind: $\Lambda\Lambda$ hypernuclei 
and $\Xi$ hyperons\thanks{Work done in collaboration with I.N. Filikhin; 
supported in part by the Israel Science Foundation, Jerusalem.}} 

\author{A. Gal\address{Racah Institute of Physics, 
The Hebrew University, Jerusalem 91904, Israel}} 
\maketitle

\begin{abstract}
Faddeev calculations suggest that: 
(i) $B_{\Lambda\Lambda}(_{\Lambda\Lambda}^{~~6}$He) for the recent KEK 
emulsion event is compatible with fairly weak $\Lambda\Lambda$ potentials 
$V_{\Lambda\Lambda}$ such as due to the Nijmegen soft-core NSC97 model; 
(ii) the isodoublet $_{\Lambda\Lambda}^{~~5}$H - $_{\Lambda\Lambda}^{~~5}$He 
hypernuclei are particle-stable even in the limit 
$V_{\Lambda\Lambda} \rightarrow 0$; and (iii) $_{\Lambda\Lambda}^{~~4}$H 
within a $\Lambda\Lambda$d model is particle-stable. 
However, four-body $\Lambda\Lambda pn$ Faddeev-Yakubovsky calculations 
do {\it not} produce a bound state for $_{\Lambda\Lambda}^{~~4}$H 
even for $V_{\Lambda\Lambda}$ considerably stronger than required to 
reproduce $B_{\Lambda\Lambda}(_{\Lambda\Lambda}^{~~6}$He), 
in contrast to the normal situation (e.g. $_{\Lambda\Lambda}^{~10}$Be) 
where a four-body Faddeev-Yakubovsky calculation yields stronger binding 
than that due to a suitably defined three-body Faddeev calculation. 
For stranger systems, Faddeev calculations 
using $\Lambda\Xi$ interactions which simulate model NSC97 suggest that 
$_{\Lambda\Xi}^{~~6}$He marks the onset of nuclear stability 
for $\Xi$ hyperons. 
\end{abstract}

\section{INTRODUCTION AND INPUT} 

Information on hyperon-hyperon ($YY$) interactions is not readily available 
from experiments in free space. It is almost exclusively limited to 
the study of strangeness $S$=$-2$ hypernuclear systems. 
This information is crucial for extrapolating into multi-strange hadronic 
matter \cite{SDG94}, for both finite systems and in bulk \cite{SBG00}, 
and into neutron stars \cite{SBG02}. 

The recent unambiguous 
identification of $_{\Lambda\Lambda}^{~~6}$He in the KEK 
hybrid-emulsion experiment E373 \cite{Tak01} is consistent with 
a scattering length $a_{\Lambda\Lambda}$$\sim$$-0.8$ fm 
\cite{FGa02b}, indicating a considerably weaker $\Lambda\Lambda$ 
interaction than that specified by $a_{\Lambda N}$$\sim$$-2$ fm 
\cite{RSY99} for the $\Lambda N$ interaction. 
With such a relatively weak $\Lambda\Lambda$ interaction, and since 
the three-body system $\Lambda \Lambda N$ is unbound (comparing it 
with the unbound $\Lambda nn$ system \cite{THe65}), the question of 
whether or not the onset of binding in the $S$=$-2$ hadronic 
sector occurs at $A$=$4$ becomes highly topical. 

Here I review primarily a recent Faddeev-Yakubovsky 
calculation for $_{\Lambda\Lambda}^{~~4}$H as a four-body 
$\Lambda\Lambda pn$ system \cite{FGa02c}, taking into account properly 
{\it all} possible rearrangement channels. I also briefly review 
calculations for other $\Lambda\Lambda$ hypernuclei and $\Lambda\Xi$ 
hypernuclei \cite{FGa02a}. 

\begin{figure} 
\begin{minipage}[t]{75mm} 
\epsfig{file=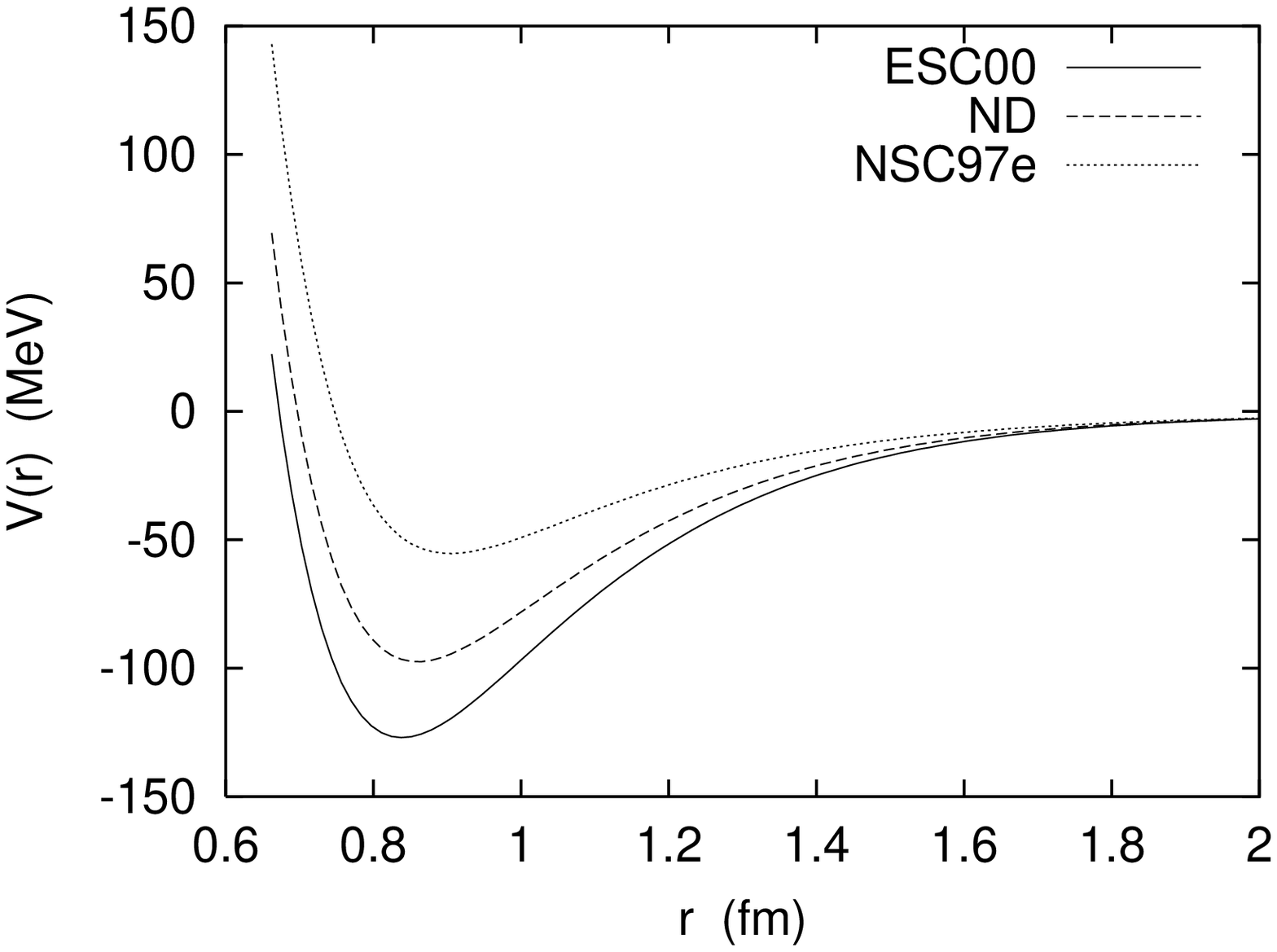,height=80mm,width=75mm} 
\caption{$\Lambda\Lambda$ potentials in OBE models.} 
\label{fig:llpot} 
\end{minipage} 
\hspace{\fill} 
\begin{minipage}[t]{75mm} 
\epsfig{file=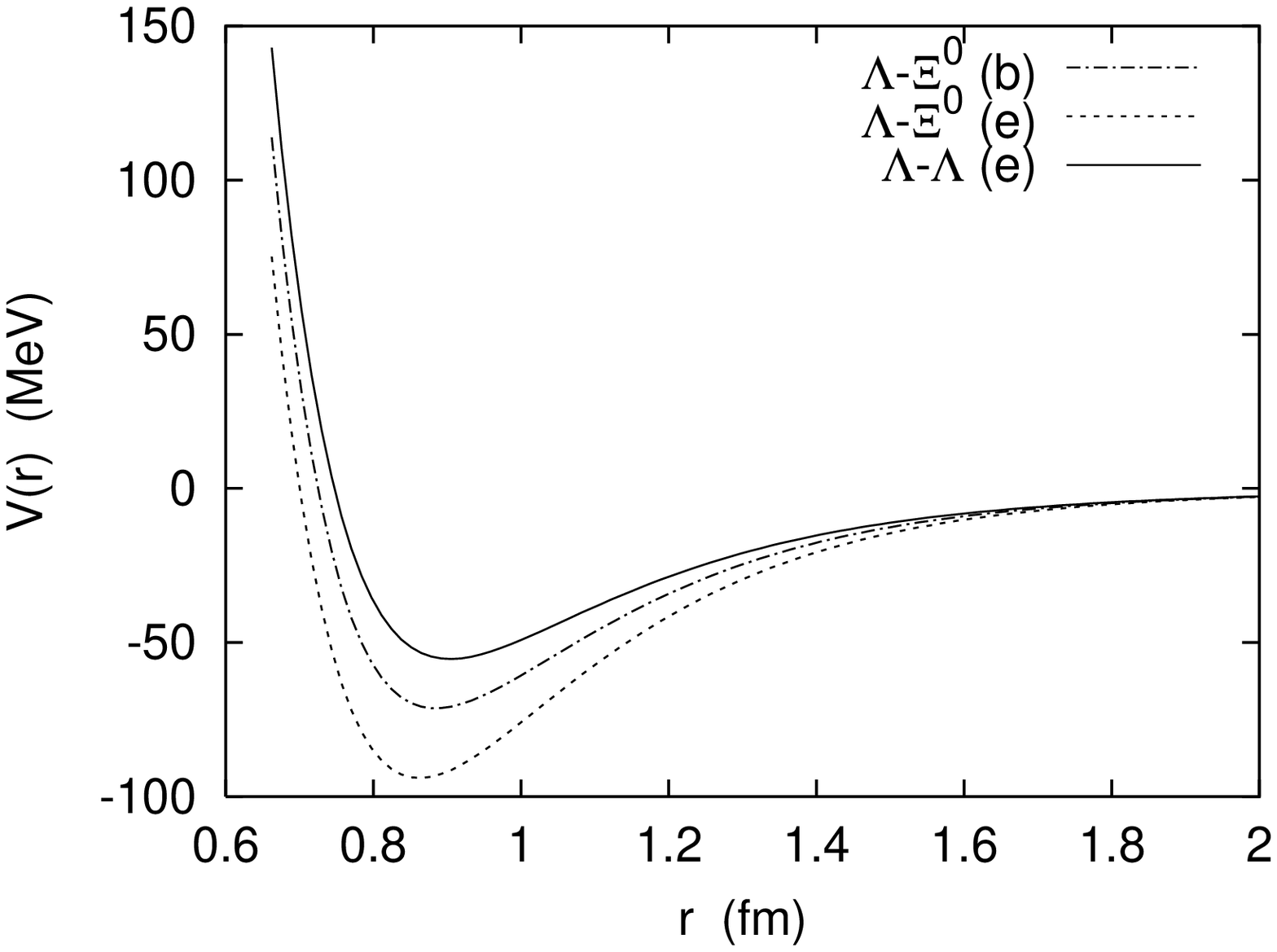,height=80mm,width=75mm} 
\caption{NSC97-model $YY$ potentials.} 
\label{fig:pot-LY} 
\end{minipage} 
\end{figure}

The $YN$ and $YY$ $s$-wave interaction input potentials consisted 
of combinations of three Gaussians with different ranges, such as those used 
by Hiyama et al. \cite{HKM97}. A similar form was used for the $pn$ 
triplet interaction and the results proved almost identical to those 
derived using the Malfliet-Tjon potential MT-III \cite{MTj69}. 
Of the several $\Lambda\Lambda$ potentials due to OBE models ahich are 
shown in Fig. \ref{fig:llpot}, NSC97e is the weakest one, of the order 
of magnitude required to reproduce 
$B_{\Lambda\Lambda}(_{\Lambda\Lambda}^{~~6}$He). The $\Lambda\Lambda$ 
interaction is fairly weak for all six versions ($a$)-($f$) of 
the Nijmegen soft-core model NSC97 \cite{SRi99}, 
and versions $e$ and $f$ provide a reasonable description of single-$\Lambda$ 
hypernuclei \cite{RSY99}. Several $YY$ potentials 
fitted to reproduce the low-energy parameters of the 
NSC97 model \cite{SRi99} are shown in Fig. \ref{fig:pot-LY}. 
We note that the $\Lambda\Xi$ interaction is rather strong, considerably 
stronger within the same version of the model (here $e$) than the 
$\Lambda\Lambda$ interaction.

\section{RESULTS} 

Figure \ref{fig:e6e5} demonstrates a nearly linear correlation between 
Faddeev-calculated values of 
$\Delta B_{\Lambda\Lambda}(_{\Lambda\Lambda}^{~~6}$He) and 
$\Delta B_{\Lambda\Lambda}(_{\Lambda\Lambda}^{~~5}$H, 
$_{\Lambda\Lambda}^{~~5}$He), using several $\Lambda\Lambda$ interactions 
including (the lowest-left point) $V_{\Lambda\Lambda} = 0$. 
The $\Lambda \Lambda$ incremental binding energy $\Delta B_{\Lambda\Lambda}$ 
is defined by 
\begin{equation} 
\label{eq:delB} 
\Delta B_{\Lambda\Lambda} (^{~A}_{\Lambda \Lambda}Z) 
= B_{\Lambda\Lambda} (^{~A}_{\Lambda \Lambda}Z) 
- 2{\bar B}_{\Lambda} (^{(A-1)}_{~~\Lambda}Z)\;, 
\end{equation} 
where $B_{\Lambda\Lambda} (^{~A}_{\Lambda \Lambda}Z)$ is the $\Lambda\Lambda$ 
binding energy of the hypernucleus $^{~A}_{\Lambda \Lambda}Z$ and 
${\bar B}_{\Lambda} (^{(A-1)}_{~~\Lambda}Z)$ is the (2J+1) average of 
$B_{\Lambda}$ values for the $^{(A-1)}_{~~\Lambda}Z$ hypernuclear core levels. 
$\Delta B_{\Lambda\Lambda}$ increases monotonically with the strength of 
$V_{\Lambda\Lambda}$, starting in approximately zero as 
$V_{\Lambda\Lambda} \rightarrow 0$, which is a general 
feature of three-body models such as the $\alpha\Lambda\Lambda$, 
$^3$H$\Lambda\Lambda$ and $^3$He$\Lambda\Lambda$ models used in these 
$s$-wave Faddeev calculations \cite{FGa02b}. The $I = 1/2$ 
$^{~~5}_{\Lambda\Lambda}$H - $^{~~5}_{\Lambda\Lambda}$He hypernuclei are 
then found to be particle stable for all the $\Lambda\Lambda$ attractive 
potentials here used. 

\subsection{$^{~~4}_{\Lambda\Lambda}$H in a $\Lambda\Lambda d$ model} 

\begin{figure} 
\begin{minipage}[t]{75mm} 
\epsfig{file=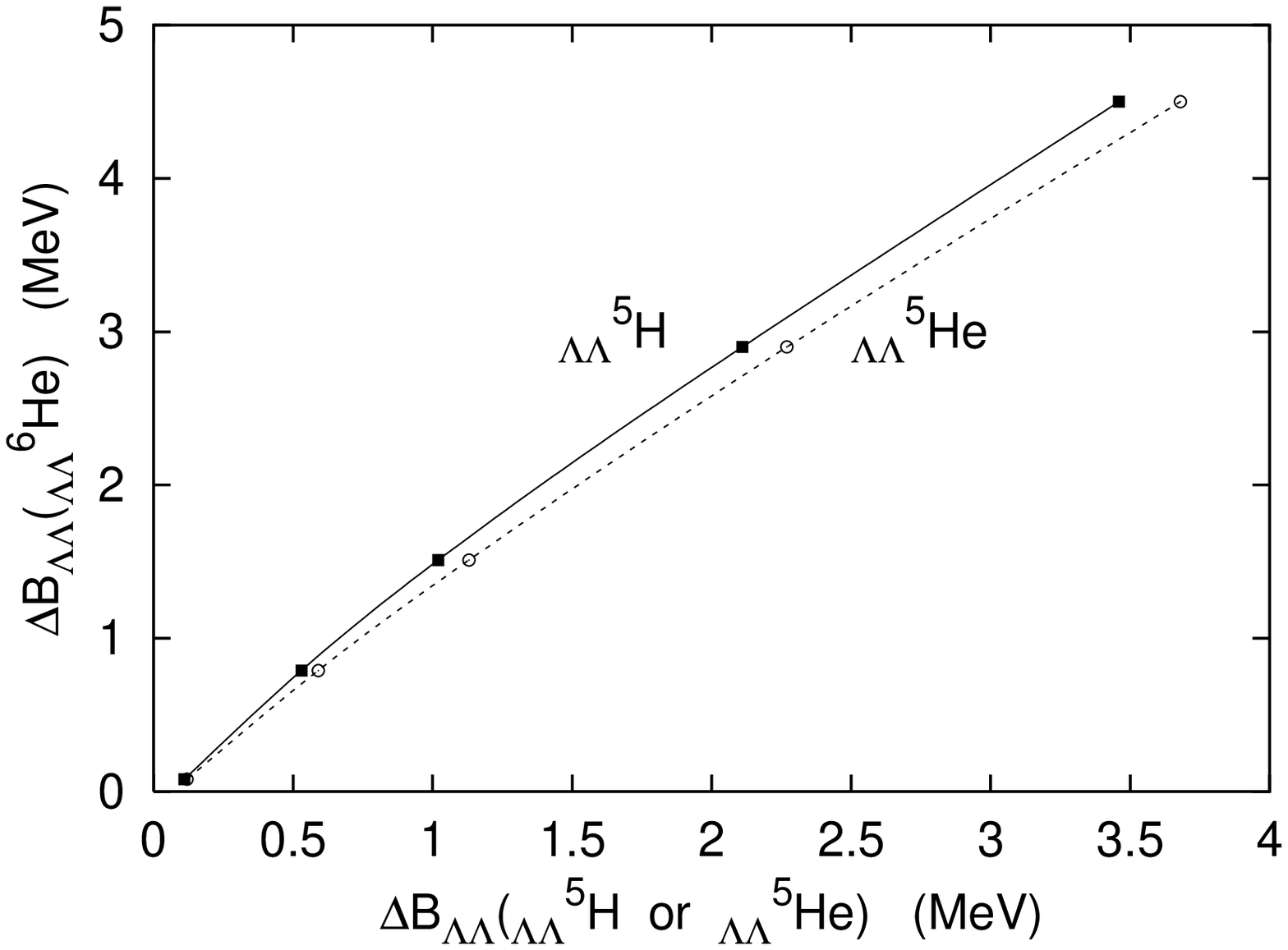,height=85mm,width=75mm} 
\caption{$s$-wave Faddeev calculations of 
$\Delta B_{\Lambda\Lambda}(_{\Lambda\Lambda}^{~~6}\rm He)$ 
vs. $\Delta B_{\Lambda\Lambda}(_{\Lambda\Lambda}^{~~5}\rm H,~ 
_{\Lambda\Lambda}^{~~5}\rm He)$.} 
\label{fig:e6e5} 
\end{minipage} 
\hspace{\fill} 
\begin{minipage}[t]{75mm} 
\epsfig{file=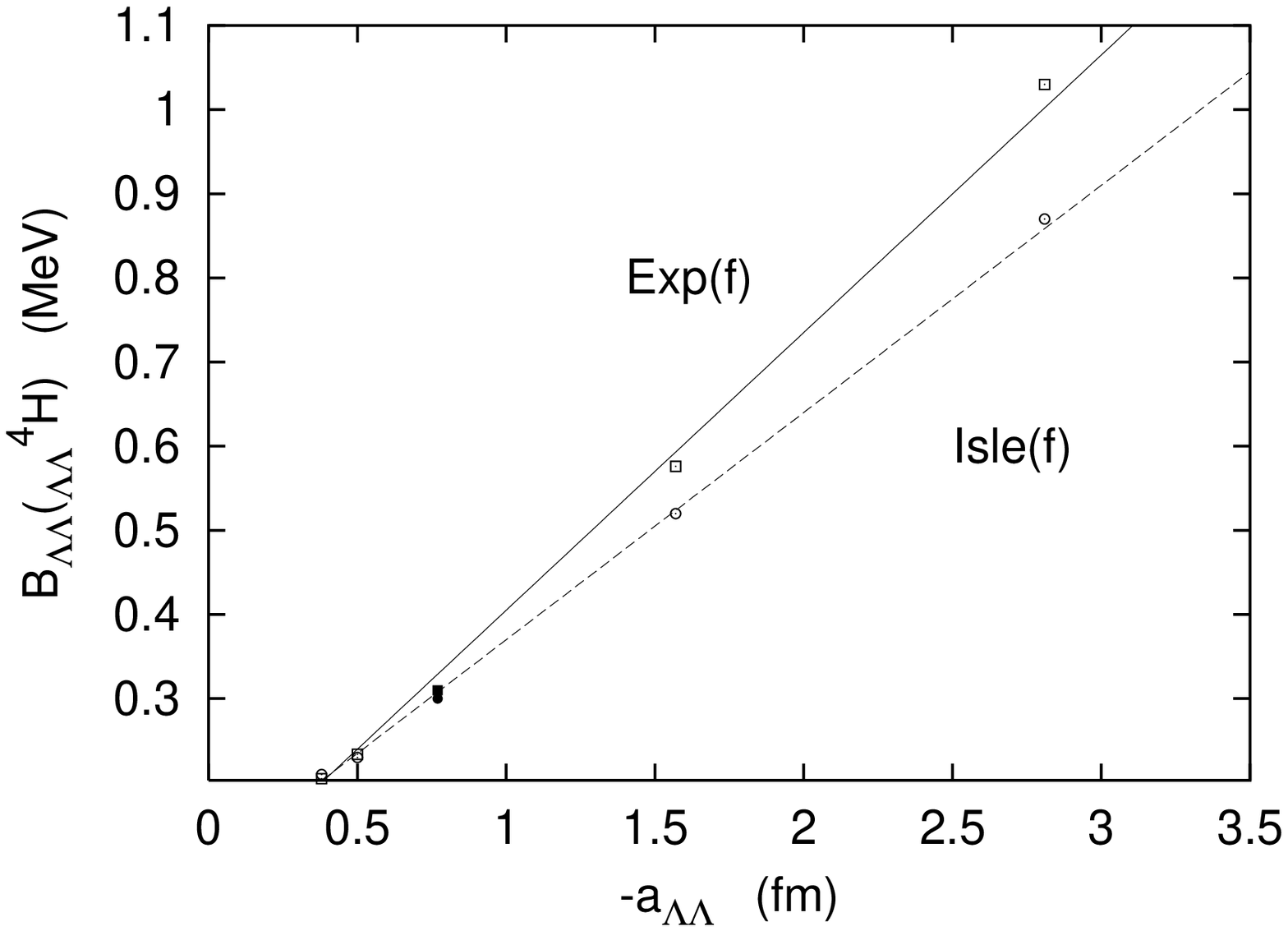,height=85mm,width=75mm} 
\caption{$s$-wave Faddeev calculations of 
$B_{\Lambda\Lambda}(_{\Lambda\Lambda}^{~~4}\rm H)$ 
vs. the scattering length $a_{\Lambda\Lambda}$.}
\label{fig:Islef} 
\end{minipage} 
\end{figure}

$\Lambda\Lambda$ binding-energy values ($B_{\Lambda\Lambda}$) calculated 
within a $\Lambda\Lambda d$ $s$-wave Faddeev calculation \cite{FGa02c} 
are shown in Fig. \ref{fig:Islef} as function of the $\Lambda\Lambda$ 
scattering length $a_{\Lambda\Lambda}$ 
for two $\Lambda d$ potentials, one of an exponential shape and the other 
one of an isle shape, both of them fitted to the low-energy parameters 
of an $s$-wave Faddeev calculation for $\Lambda pn$ which uses model NSC97f 
for the underlying $\Lambda N$ interaction, yielding 
$B_{\Lambda}(_{\Lambda}^3{\rm H}({\frac{1}{2}}^+))=0.19$ MeV.\footnote
{Using model NSC97e, with $B_{\Lambda}(_{\Lambda}^3{\rm H})=0.07$ MeV, 
does not alter the conclusions listed below.} 

The roughly linear increase of $B_{\Lambda\Lambda}$ with the strength of 
$V_{\Lambda\Lambda}$ holds generally \cite{FGa02b} in three-body 
$\Lambda \Lambda C$ models ($C$ standing for a cluster) as is evident also 
from Fig. \ref{fig:e6e5}. The solid squares in Fig. \ref{fig:Islef} 
correspond to a $\Lambda\Lambda$ interaction 
fitted to $B_{\Lambda\Lambda}(_{\Lambda\Lambda}^{~~6}\rm He)$. 
The onset of particle stability for 
$^{~~4}_{\Lambda\Lambda}$H($1^+$) requires a minimum strength for 
$V_{\Lambda\Lambda}$ which is exceeded by the choice 
of $B_{\Lambda\Lambda}(^{~~6}_{\Lambda\Lambda}$He) \cite{Tak01} 
as a normalizing datum. 
Disregarding inessential complications due to spin it can be shown that, 
for essentially attractive $\Lambda\Lambda$ interactions and 
for a static nuclear core $d$, a two-body $\Lambda d$ bound state implies 
binding for the three-body $\Lambda\Lambda d$ system \cite{Bas02}. 

\subsection{$^{~~4}_{\Lambda\Lambda}$H in a $\Lambda\Lambda pn$ model} 

For two identical hyperons and two essentially identical nucleons 
(upon introducing isospin) as appropriate to a $\Lambda\Lambda pn$ 
model calculation  of $^{~~4}_{\Lambda\Lambda}$H, the 18 Faddeev-Yakubovsky 
components reduce to seven independent components satisfying coupled 
equations. Six rearrangement channels are involved in the $s$-wave 
calculation 
\cite{FGa02c} for $^{~~4}_{\Lambda\Lambda}$H(1$^+$): 
\begin{equation} 
(\Lambda NN)_{S=\frac12} + \Lambda \;, \;\; 
(\Lambda NN)_{S=\frac32} + \Lambda \;, \;\; 
(\Lambda \Lambda N)_{S=\frac12} + N \; 
\label{eq:3+1} 
\end{equation} 
for 3+1 breakup clusters, and 
\begin{equation} 
(\Lambda \Lambda)_{S=0} + (NN)_{S=1} \;, \;\;\; 
(\Lambda N)_S + (\Lambda N)_{S'} 
\label{eq:2+2} 
\end{equation} 
with $(S,S')$=$(0,1)$+$(1,0)$ and $(1,1)$ for 2+2 breakup clusters. 

\begin{figure} 
\begin{minipage}[t]{75mm} 
\epsfig{file=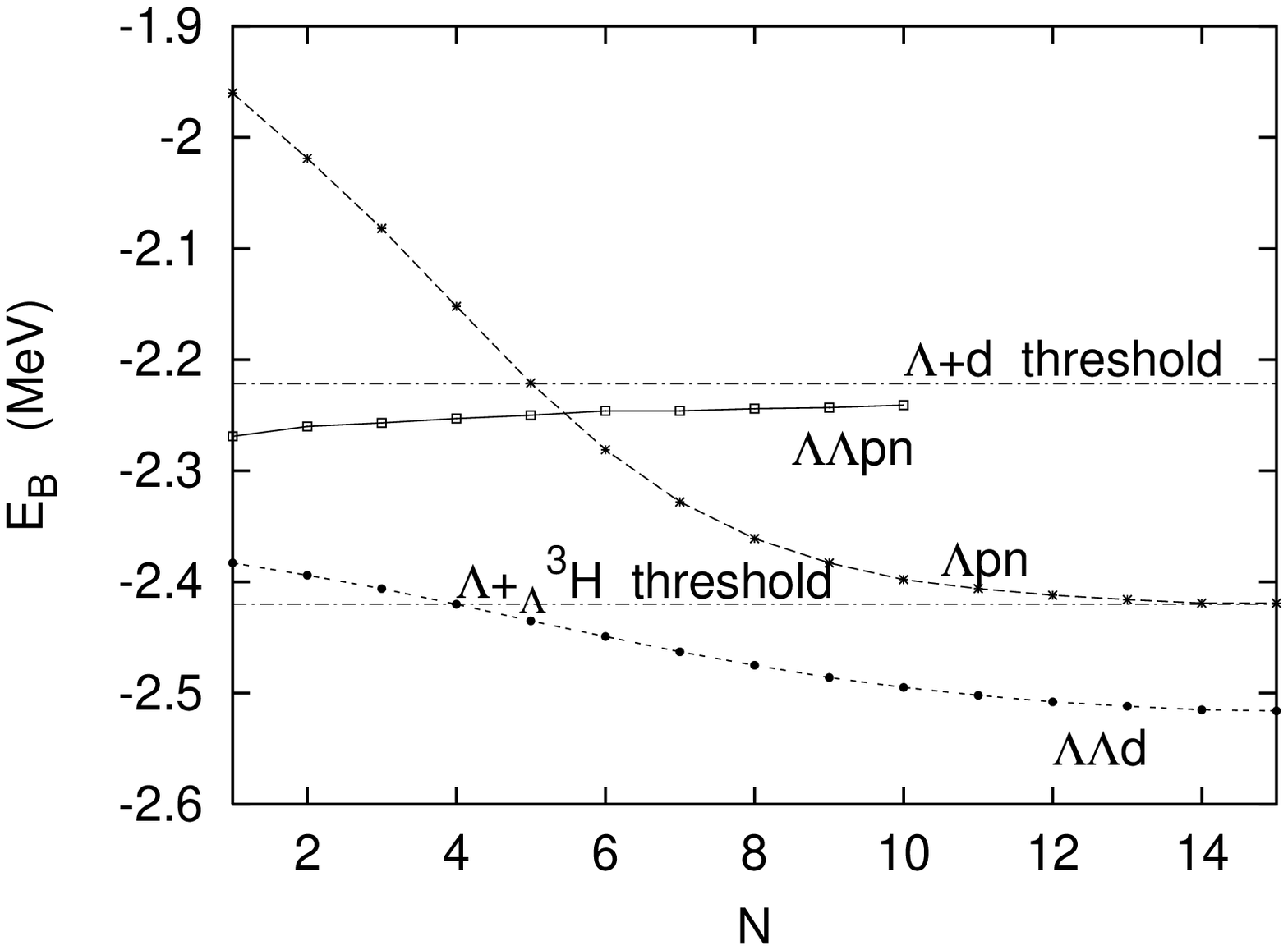,height=85mm,width=75mm} 
\caption{$s$-wave Faddeev-Yakubovsky calculations for 
$\Lambda pn$, $\Lambda\Lambda d$ and $\Lambda\Lambda pn$.} 
\label{fig:BE} 
\end{minipage} 
\hspace{\fill} 
\begin{minipage}[t]{75mm} 
\epsfig{file=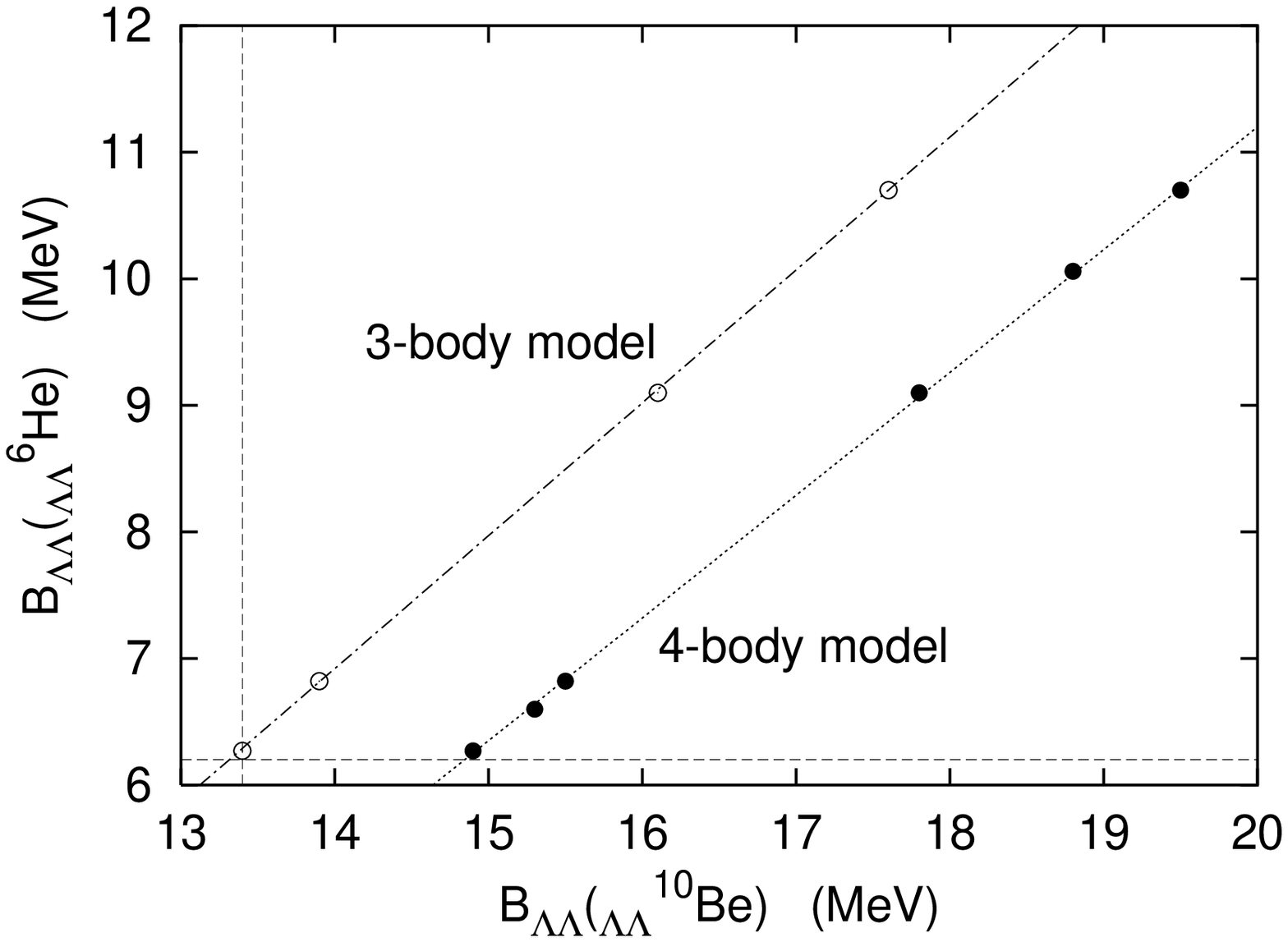,height=85mm,width=75mm} 
\caption{$s$-wave Faddeev-Yakubovsky calculations for 
$^{~10}_{\Lambda\Lambda}$Be: $^8$Be$\Lambda \Lambda$ vs. 
$\alpha\alpha\Lambda\Lambda$.} 
\label{fig:e6e10} 
\end{minipage} 
\end{figure}

Using $V_{\Lambda\Lambda}$ which reproduces 
$B_{\Lambda\Lambda}(^{~~6}_{\Lambda\Lambda}$He), 
the four-body calculation converges well as function of the number 
$N$ of the Faddeev-Yakubovsky basis functions allowed in, 
yet yielding no bound state for the $\Lambda\Lambda pn$ 
system, as demonstrated in Fig. \ref{fig:BE} by the location of the 
`$\Lambda\Lambda pn$' curve {\it above} the horizontal straight line 
marking the `$\Lambda + ~_{\Lambda}^{3}{\rm H}$ threshold'.\footnote{This 
threshold was obtained as the asymptote of the $\Lambda pn$ 
$s$-wave Faddeev calculation; note that, in agreement with 
Fig. \ref{fig:Islef}, the asymptote of the `$\Lambda\Lambda d$' curve 
is located {\it below} this threshold.} 
In fact these Faddeev-Yakubovsky calculations exhibit little sensitivity 
to $V_{\Lambda\Lambda}$ over a wide range. 
Even for considerably stronger $\Lambda\Lambda$ interactions one gets a bound 
$^{~~4}_{\Lambda\Lambda}$H only if the $\Lambda N$ interaction is made 
considerably stronger, by as much as 40\%. With four $\Lambda N$ pairwise 
interactions out of a total of six, the strength of 
the $\Lambda N$ interaction (about half of that for $NN$) plays a major 
role in the four-body $\Lambda\Lambda pn$ problem. 
Here it appears impossible to prove that, for essentially attractive 
$\Lambda\Lambda$ interactions and for a {\it non static} nuclear core $d$ 
(made out of dynamically interacting proton and neutron), 
a $\Lambda d$ bound state implies binding for the $\Lambda\Lambda d$ system. 
It is unlikely that incorporating higher partial waves, and 
$\Lambda\Lambda - \Xi N$ coupling effects, will change this qualitative 
feature.

\subsection{$^{~10}_{\Lambda\Lambda}$Be} 

For heavier $\Lambda\Lambda$ hypernuclei, 
the relationship between the three-body and four-body models 
is opposite to that found here for $^{~~4}_{\Lambda\Lambda}$H: 
the $\Lambda\Lambda C_1 C_2$ calculation provides {\it higher} binding 
than a properly defined $\Lambda\Lambda C$ calculation yields 
(with $C=C_1+C_2$) due to the attraction induced by 
the $\Lambda C_1$-$\Lambda C_2$, $\Lambda\Lambda C_1$-$C_2$, 
$C_1$-$\Lambda\Lambda C_2$ four-body rearrangement channels that include 
bound states for which there is no room in the three-body $\Lambda\Lambda C$ 
model. The binding energy calculated within the four-body model increases 
then `normally' with the strength of $V_{\Lambda\Lambda}$ \cite{FGa02b}. 
This is demonstrated in Fig. \ref{fig:e6e10} for $^{~10}_{\Lambda\Lambda}$Be 
using several $\Lambda\Lambda$ interactions, including $V_{\Lambda\Lambda}=0$ 
which corresponds to the lowest point on each one of the straight lines. 
The origin of the dashed axes corresponds to $\Delta B_{\Lambda\Lambda} = 0$. 
The fairly large value of 1.5 MeV for 
$\Delta B_{\Lambda\Lambda}(^{~10}_{\Lambda\Lambda}$Be) in the limit 
$V_{\Lambda\Lambda} \rightarrow 0$ is due to the special $\alpha\alpha$ 
cluster structure of the $^8$Be core. The correlation noted in the figure 
between $^{~10}_{\Lambda\Lambda}$Be and $^{~~6}_{\Lambda\Lambda}$He 
calculations, and the consistency between various reports on their 
$B_{\Lambda\Lambda}$ values, are discussed in Ref. \cite{FGa02b} 
and by Hiyama in these proceedings. 

\subsection{The onset of $\Xi$ stability} 

\begin{figure}[hbt] 
\epsfig{file=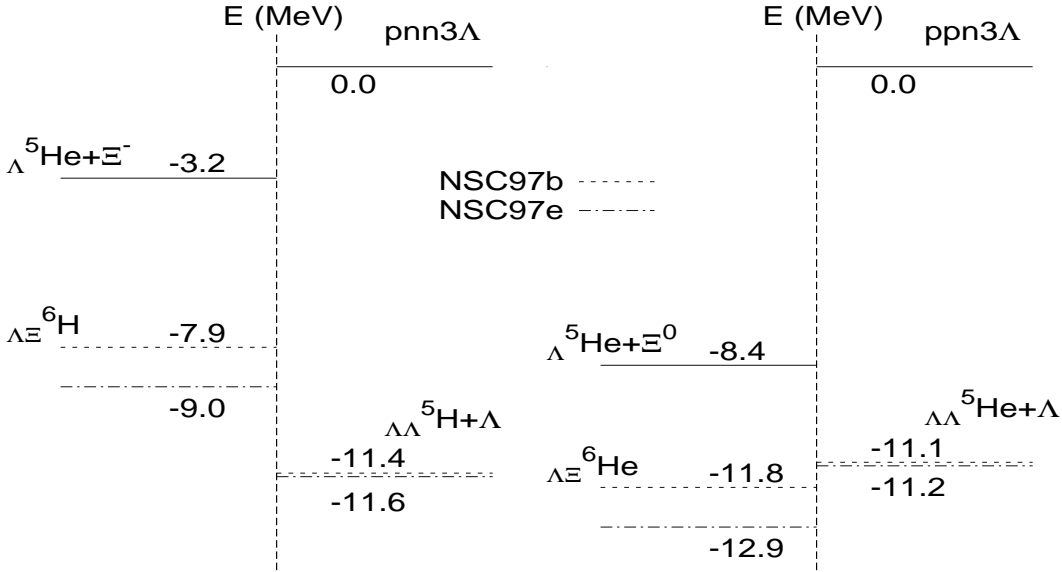,height=75mm,width=140mm} 
\caption{$s$-wave Faddeev calculations for the level schemes of 
$_{\Lambda\Xi}^{~~6}$H and $_{\Lambda\Xi}^{~~6}$He.} 
\label{fig:LXi} 
\end{figure} 

If model NSC97 indeed provides a valid extrapolation from fits 
to $NN$ and $YN$ data, and recalling the strongly attractive 
$^{1}S_{0}$ $\Lambda\Xi$ potentials due to this model (Fig. \ref{fig:pot-LY}), 
it is only natural to search for stability of $A=6, S=-3$ systems 
obtained from $^{~~6}_{\Lambda\Lambda}$He replacing one of the 
$\Lambda$'s by $\Xi$. Faddeev calculations \cite{FGa02a} 
for the isodoublet hypernuclei $^{~~6}_{\Lambda\Xi}$H and 
$^{~~6}_{\Lambda\Xi}$He, considered as $\alpha\Lambda\Xi^-$ and 
$\alpha\Lambda\Xi^0$ three-body systems respectively, 
indicate that $^{~~6}_{\Lambda\Xi}$He is particle-stable 
against $\Lambda$ emission to $^{~~5}_{\Lambda\Lambda}$He 
for potentials simulating model NSC97, 
particularly versions $e$ and $f$, 
whereas $_{\Lambda\Xi}^{~~6}$H is unstable since 
$M(\Xi^-) > M(\Xi^0)$ by 6.5 MeV.\footnote{Recall that the $I=1/2$ 
$^{~~5}_{\Lambda\Lambda}$H - $^{~~5}_{\Lambda\Lambda}$He 
hypernuclei, within a $\Lambda\Lambda C$ Faddeev calculation, 
are particle stable even in the limit $V_{\Lambda\Lambda} \rightarrow 0$.} 
This is demonstrated in Fig. \ref{fig:LXi}. Nevertheless, predicting 
particle stability for $^{~~6}_{\Lambda\Xi}$He is not independent 
of the assumptions made on the experimentally unexplored $\Xi \alpha$ 
interaction which was extrapolated from recent data on $^{12}$C \cite{Fuk98}; 
hence this prediction cannot be considered conclusive.

\section{CONCLUSION} 

I have presented a first ever four-body Faddeev-Yakubovsky calculation 
\cite{FGa02c} for $^{~~4}_{\Lambda\Lambda}$H using $NN$ and $\Lambda N$ 
interaction potentials that fit the available data on the relevant 
subsystems, including the binding energy of $^3_\Lambda$H. 
No $^{~~4}_{\Lambda\Lambda}$H bound state was obtained for 
a wide range of $\Lambda\Lambda$ interactions, including that corresponding 
to $B_{\Lambda\Lambda}(^{~~6}_{\Lambda\Lambda}$He). 
This non binding is due to the relatively weak $\Lambda N$ 
interaction, in stark contrast to the results of 
a `reasonable' three-body $\Lambda\Lambda d$ Faddeev 
calculation. More experimental work 
is needed to decide whether or not the events reported recently in the 
AGS experiment E906 \cite{Ahn01} correspond to $^{~~4}_{\Lambda\Lambda}$H. 

Accepting the predictive power of model NSC97, 
our calculations suggest that $_{\Lambda\Xi}^{~~6}$He 
may be the lightest particle-stable $S=-3$ hypernucleus, 
and the lightest and least strange particle-stable hypernucleus 
in which a $\Xi$ hyperon is bound (and not $_{\Lambda\Lambda\Xi}^{~~~7}$He 
for $S=-4$ as argued in Ref. \cite{SDG94}). Unfortunately, the direct 
production of $\Lambda\Xi$ hypernuclei is beyond 
present experimental capabilities, requiring the use of $\Omega^-$ 
initiated reactions.

\section*{ACKNOWLEDGMENTS} 

Special thanks are due to my collaborator Igor Filikhin who has devotedly 
and patiently introduced me to modern aspects of Few-Body Physics, 
a research field I have not attended since the pioneering $K^- d$ reactive 
Faddeev calculation \cite{TGE79}.

\end{document}